\documentclass[11pt,a4paper]{article}
\begin{document}

\title{ Does the Schwarzschild black hole really exist? }
\author{Guihua Tian$^{1,2}$\\
1.School of Science, Beijing University \\
of Posts And Telecommunications. Beijing, 100876, China.\\2.KLMM,
AMSS, Chinese Academy of Sciences(CAS), Beijing, 100080, China}
\date{}
\maketitle

\begin{abstract}
we use the Kruskal time coordinate $T$ to define the initial time.
By this way, it naturally divides the stable study  into one
connected with the two regions: the white-hole-connected region
and the black-hole-connected region. The union of the two regions
covers the Schwarzschild space-time ($r\ge 2m$). We also obtain
the very reasonable conclusion: the white-hole-connected region is
instable; whereas the black-hole-connected region is stable. If we
take the instability with caution and seriousness, it might be not
unreasonable to regard that the Schwarzschild black hole might be
instable too.

\textbf{PACC:0420-q}
\end{abstract}

In general relativity, black holes could be produced by the
complete gravitational collapse of sufficiently massive stars.
Whether or not the black holes exist depends on their stability
against small perturbation. The Schwarzschild black hole is the
only vacuum spherically symmetric solution of the Einstein's
equation. It is the final fate of a spherical star after its
complete gravitational collapse. Of course, the perturbation study
to the Schwarzschild black-hole is of critical importance due to
its very special status in general relativity. It is treated and
studied by many people since Regge-Wheeler's original paper in
1957\cite{rw}. Generally, the problem of the stability of the
Schwarzschild black hole is believed to be settled
down\cite{vish}-\cite{wald}: it is absolutely stable against small
perturbation.

Here we study the problem again. Due to its metric component
$g_{00}=0$ at the horizon $r=2m$, the Schwarzschild time
coordinate $t$ lose its meaning at the horizon. This is the
drawback of the Schwarzschild time coordinate $t$. Actually, this
drawback is closely connected with the boundary problem at the
horizon $r=2m$ in the stability study. People have not consider
its effect in all previous stability study\cite{rw}-\cite{wald}:
$t=0$ is taken for granted as the initial time at the horizon. It
is natural for one to suspect its rationality. One might guess its
influence on the stability study. It would best for one to use
some "good" time to reconsider the stability problem and compare
the two results. It is the main work in this paper.

By using the more reasonable and suitable time coordinate, the
Kruskal time coordinate $T$, to define the initial time, we study
the stable problem and obtain the conclusion completely opposite
to the common belief: the Schwarzschild black hole might be
instable from some viewpoint (see the text). This unexpected
conclusion would urge people to study the perturbation of the Kerr
black hole, which is the second simplest black hole and
topologically different from the Schwarzschild black hole. More
importantly, the Kerr black hole has intrinsic angular momentum
produced more possible from the spinning star's complete
gravitational collapse.

In Regge and Wheeler's original paper,  they deduced perturbation
into two kinds\cite{rw}: the odd one and the even one,
corresponding respectively to the angular and radial metric
perturbation\cite{chan}. In the paper, we study the odd
perturbation.

For the Schwarzschild black hole, the perturbation equations were
obtained in the Schwarzschild coordinates:
\begin{equation}
ds^{2}=-(1-\frac{2m}{r})dt^{2}+(1-\frac{2m}{r})^{-1}dr^{2}+r^{2} d
\Omega ^{2},\label{orimetric}
\end{equation}
and the odd perturbation equation is the well-known Regge-Wheeler
equation\cite{rw}:
\begin{equation}
\frac{d^{2}Q}{dr^{*2}}+\left[k^{2}-V\right]Q=0\label{reggewheeler},
\end{equation}
where the tortoise coordinate $r^*$ and the potential $V$ are
defined as
\begin{equation}
r^{*}=r+2m\ln(\frac{r}{2m}-1),
\end{equation}
\begin{equation}
V=(1-\frac{2m}{r})[\frac{l(l+1)}{r^{2}}-\frac{6m}{r^{3}}].
\end{equation}
The function $Q$ is connected with the odd perturbation fields
$h^s_{\mu \nu }$ (the upper-case index s denote the corresponding
quantity is in the Schwarzschild coordinates) by
\begin{equation}
h_{13}^{s}=r\left(1-\frac{2m}{r}\right)^{-1}Qe^{-ikt}
\end{equation}
and
\begin{equation}
h_{03}^{s}=\frac{i}{k}\left[\left(1-\frac{2m}{r}\right)Q
+r\frac{dQ}{dr^{*}}\right]e^{-ikt}.\label{h0white}
\end{equation}

The boundary for the odd perturbation equation
(\ref{reggewheeler}) consists of the spatial infinity and the
horizon $r=2m$ of the black hole. The spatial infinity boundary is
easy to treat because the background metric is regular there. On
the contrary, it is very difficulty to treat the boundary at the
horizon $r=2m$. The background metric $g^s_{\mu \nu}$ has an
apparent singularity at $r=2m$ in the Schwarzschild coordinates.
It is difficult to distinguish the real divergence in perturbation
fields from the spurious one caused by the background metric's
ill-defined-ness at $r=2m$. This is the first problem connected
with the boundary at $r=2m$. Vishveshwara first noticed this
problem and proposed a solution to it\cite{vish}.

Not only does the background metric has an apparent singularity at
$r=2m$ in the Schwarzschild coordinates, but also does its time
coordinate $t$  lose its time meaning$^5$ at $r=2m$. For example:
a traveller was dropped into the black hole to across the horizon
$r=2m$, and he only need to take infinitesimal proper time to
cross the horizon. But the process will take infinite long "time"
if we use the coordinate $t$ to analyze it. This consists of the
second problem at boundary of the horizon of the black hole. The
second problem is much more elusive, and still escapes from the
people's notices.

The discovery of the Kruskal coordinates may send some lights on
these two problems. In Kruskal coordinates, the background metric
is
\begin{equation}
ds^{2}=\frac{32m^3}{r}e^{-\frac
r{2m}}\left[-dT^{2}+dX^{2}\right]+r^{2} d \Omega
^{2}.\label{kruskalmetric}
\end{equation}
In addition, the relations between $(t,\ r)$ and $(T,\ X)$ are
\begin{equation} (\frac
r{2m}-1)e^{-\frac r{2m}}=X^2-T^2\label{relation tT1}
\end{equation}
and
\begin{equation}
e^{\frac t{4m}}=\frac {X+T}{X-T}.\label{relation tT2}
\end{equation}
The metric (\ref{kruskalmetric})is regular at $r=2m$. More
importantly, the Kruskal time coordinate $T$ is more physically
reasonable to analyze the process cross or near or at the surface
$r=2m$ of the horizon\cite{mtw}. It is best to study the stable
problem completely in Kruskal coordinates. Nevertheless, it is
generally believed that this is too difficulty to be finished due
to its non-stationary property of the background in Kruskal
coordinates system.

Vishveshwara proposed an easier solution to the first problem. The
background metric is well-behaved (or in other word, it is
asymptotically flat) at the spatial infinity in the Schwarzschild
coordinates. Therefore, He first selected solutions falling off to
zero as $r\rightarrow \infty$ in the Schwarzschild coordinates,
and obtained their asymptotic forms near the horizon $r=2m$. Then,
he used the Kruskal coordinates system only to study critically
the perturbation fields near the horizon $r=2m$. In this way,
Vishveshwara solved the first problem connected with the boundary
at the horizon $r=2m$. But in his final step, he defined the
Schwarzschild time coordinate $t=0$ to be the initial time, and
obtained that the Schwarzschild black hole is stable.

Actually, the Schwarzschild time coordinate $t$  loses its time
meaning at $r=2m$\cite{mtw}. We want to know whether or not this
drawback of its time coordinate $t$ have effect and influence in
the study of the stability problem of the black hole. To obtain
definite answer to it, one might try to use other "good" time to
study the stability problem and compare the results with that of
Vishveshwara's.

Following Vishveshwara, select the well-behaved solutions at
spatial infinity as those that fall off to zero as $r\rightarrow
\infty $. For pure positive imaginary frequency $k=i\alpha $ with
$\alpha > 0$, the asymptotic form of these good solutions is
\begin{equation}
Q_{\infty}=\tilde{A} e^{ikr^{*}}\label{q-infinty}
\end{equation}
where the function satisfies the Regge-Wheeler equation
(\ref{reggewheeler}), which determines the evolution of the
function $Q$ from $r\rightarrow \infty $ to $r=2m$. By
eq.(\ref{reggewheeler}), the asymptotic form of these good
solutions must be
\begin{equation}
Q_{2m}=Ae^{ikr^{*}}\label{q-2m}
\end{equation}
at $r=2m$ \cite{vish}. The first problem connected with the
boundary at $r=2m$ in the Schwarzschild coordinates made
Vishveshwara transform $h^s_{\mu \nu }$ into $h^k_{\mu \nu }$ of
the Kruskal forms as\cite{vish}
\begin{equation}
h_{03}^{k}=-8m^2A\left(X-T\right)^{-4m\alpha-1} \label{h03-k}
\end{equation}
and
\begin{equation}
h_{13}^{k}=8m^2A\left(X-T\right)^{-4m\alpha-1} \label{h13-k}
\end{equation}
near $r=2m$ with $k=i\alpha $. Here we must decide to select what
time coordinate to define the initial time.

As we state before, Vishveshwara selected the time $t=0$ to be the
initial time and obtained that the Schwarzschild black hole cannot
have pure positive imaginary frequency. In this way, he proved the
Schwarzschild black hole is stable.

We now select the Kruskal time $T$ as "good" time to define the
initial time at $r=2m$. If we regard the Schwarzschild black hole
is formed by star's collapse in the remote past, it is reasonable
to study the stable problem by $T<0$. We could choose $T=c_1<0$ as
the initial time for the boundary at $r=2m$. In this case, $r=2m$
corresponds $X+T=0$, so initially $X=-T=-c_1$ at $r=2m$. By
eqs.(\ref{h03-k}), (\ref{h13-k}), it is easy to see that the
perturbation fields $h^k_{13},\ h^k_{03}$ all are well-behaved at
$T=c_1$ and $X+T=0$. As the time grows, the perturbation fields
$h^k_{13},\ h^k_{03}$
\begin{equation}
h_{03}^{k}=-8m^2A\left(-2T\right)^{-4m\alpha-1}\label{h03-kwhite},
\end{equation}
\begin{equation}
h_{13}^{k}=8m^2A\left(-2T\right)^{-4m\alpha-1}\label{h13-kwhite}.
\end{equation}
all grows too. Finally, They blows up into infinity near $T=0$.
Therefore, the black hole might be regard as instable with
perturbation. Of course, one must notices the following facts:
this instability is closely connected with the part $T<0,\ X\ge
0,\ X^2-T^2\geq 0$ corresponding the Schwarzschild region $r\geq
2m$ connected with the white-hole, which is now defined as the
white-hole-connected region. Therefore, the white-hole-connected
region is actually instable with the perturbation.

Alternatively , we could also choose  $T=c_2\geq 0$ as the initial
time for the boundary at $r=2m$ for the black hole. In the case,
$r=2m$ corresponds $X-T=0$, so initially $X=T=c_2$. By
eqs.(\ref{h03-k}), (\ref{h13-k}), it is easy to see that the
perturbation fields $h^k_{13},\ h^k_{03}$ are blows up initially
at $T=c_2$ and $X-T=0$. That is, we could make the perturbation
fields as large as we wish by $X-T\rightarrow 0$ at the initial
time $T=c_2$. This corresponds the perturbation fields initially
are not satisfied the requirement of the well-behaved-ness, so,
they are unacceptable. Therefore, the possibility of the pure
positive imaginary frequency for the perturbation is excluded and
the stability is proved in the usual physicist' view point.

Similarly, this stability is closely connected with the part $T>0,
\ X\ge 0,\ X^2-T^2\geq 0$ corresponding the Schwarzschild region
$r\geq 2m$ connected with the black-hole, which is now defined as
the black-hole-connected region. Therefore, the above proof of
stability belongs to that of the stability of the
black-hole-connected region.

In above proof, we use the Kruskal time coordinate $T$ to define
the initial time. By this way, it naturally divides the stable
study  into one connected with the two regions: the
white-hole-connected region $X^2-T^2\geq 0,\ T\leq 0,\  X\geq 0$
and the black-hole-connected region $X^2-T^2\geq 0,\ T\geq 0,\
X\geq 0$. The union of the two regions covers the Schwarzschild
space-time in the Kruskal coordinates $X^2-T^2\geq 0,\ X\geq 0$.
We also obtain the very reasonable conclusion: the
white-hole-connected region is instable in nature because it
connects with the white hole, from which every particle must leave
in a finite time; whereas the black-hole-connected region is
stable.

If we choose other "good" times like the Painlev\'{e}' time to
define the initial time, could the similar results  also be
obtained? The answer is certainly positive, see
references\cite{tian}-\cite{tian2}.

The instability of the white-hole-connected region means that only
the part $T\ge 0, X+T\ge 0$ do have physical meaning in the
Kruskal coordinates and the Schwarzschild space-time could be used
in Schwarzschild coordinates only with $t\ge 0$. But this is not
concede with the usual viewpoint. People would feel unease at this
constraint ($t\ge 0$).

This is just one viewpoint to explain the above results, there are
also other completely different viewpoints. For example, if we
take the instability of the white-hole-connected region seriously,
then it might mean the following: the Schwarzschild black hole is
formed in the remote past and subsequently should be liable to the
perturbation in $T=c_1<0$, therefore it could not exist by this
kind instability and would evolute into other kind black holes.
Consequently, the black-hole-connected region is no longer in
existence and its stability is meaningless.

If the second viewpoint is correct, it may also mean that the
black holes should be of the Kerr or other types too. Of course,
the first selection is the Kerr black hole. Therefore, the study
of the Kerr black hole becomes critically urgent and important in
general relativity.

\section*{Acknowledgments}
We are supported in part by the National Science Foundation of
China under Grant No.10475013, No.10373003, No.10375087,
No.10375008, the National Basic Research Program of China:
NKDRTC(2004CB318000) and the post-doctor foundation of China.

\end{document}